\documentclass[aps,prl,preprint,showpacs]{revtex4}
\usepackage{bm}
\usepackage{graphicx}
\begin{document}
\title{Scaling and universality of critical fluctuations in granular gases}
\author{J. Javier Brey}
\email{brey@us.es}
\author{M.I. Garc\'{i}a de Soria}
\author{P. Maynar}
\author{M.J. Ruiz-Montero}
\affiliation{F\'{\i}sica Te\'{o}rica, Universidad de Sevilla,
Apdo.\ de
 Correos 1065, E-41080 Sevilla, Spain}

\date{today}

\begin{abstract}
The global energy fluctuations of a low density gas granular gas
in the homogeneous cooling state near its clustering instability
are studied by means of molecular dynamics simulations. The
relative dispersion of the fluctuations is shown to exhibit a
power-law divergent behavior. Moreover, the probability
distribution of the fluctuations presents data collapse as the
system approaches the instability, for different values of the
inelasticity. The function describing the collapse turns out to be
the same as the one found in several molecular equilibrium and
non-equilibrium systems, except for the change in the sign of the
fluctuations.
\end{abstract}
\pacs{PACS Numbers: 45.70.-n,51.10.+y,05.20.Dd}

\maketitle

A granular system is an assembly of macroscopic particles
dissipating their energy through inelastic collisions. In the
rapid granular flow regime, the grains move freely and
independently between collisions and the system is often referred
to as a granular gas. In this regime, the behavior of a granular
system resembles that of a molecular fluid, although with many
significant differences \cite{JNyB96}. These discrepancies have
been extensively illustrated by using theoretical, experimental,
and also particle simulation methods \cite{Go03}. In this paper,
evidence will be reported of a rather surprising analogy between
granular gases and molecular systems. It will be shown that the
statistical behavior of a freely evolving granular system near the
clustering instability has strong similarities with that of
several molecular model systems, including equilibrium systems in
the critical region. More precisely, the second moment of the
relative fluctuations of the global energy diverges with a given
critical exponent. Besides, and even more remarkably, the energy
fluctuations obey a scaling law that, when properly expressed, is
the same as the one found in some correlated equilibrium and
non-equilibrium molecular systems \cite{BHyP98,BChFyal00,AyG01}.

A widely used simple model for granular gases is a system of
smooth inelastic hard spheres ($d=3$) or disks ($d=2$) of mass $m$
and diameter $\sigma$. The inelasticity of collisions is
characterized by a constant coefficient of normal restitution
$\alpha$. As a consequence of the energy dissipation in
collisions, isolated granular gases do not exhibit any homogeneous
time-independent state. The simplest state is the so-called {\em
homogeneous cooling state} (HCS) \cite{Ha83} with vanishing flow
field and a monotonically decreasing temperature $T_{hcs}(t)$,
obeying the equation $\partial_{t}
T_{hcs}(t)=-\zeta_{hcs}(T_{hcs})T_{hcs}(t)$, where
$\zeta_{hcs}(T_{hcs}) \propto T_{hcs}^{1/2}$ is the cooling rate.
At a microscopic level, it has been postulated that the time
dependence of the ensemble describing this state occurs only
through the scaling of the velocities with the thermal velocity,
that is proportional to $T_{hcs}^{1/2}(t)$, and the corresponding
normalization \cite{GyS95}. The HCS is unstable against long
wavelength spatial perturbations \cite{GyZ93}, leading to the
spontaneous formation of velocity vortices and density clusters.
This clustering instability is accurately predicted by a linear
stability analysis of the hydrodynamic equations, that shows that
it is driven by the transversal shear mode \cite{GyZ93,BRyC99}. A
critical length $L_{c}$ is identified, so that the system becomes
unstable when its linear size $L$ exceeds $L_{c}$. Its value has
been determined for a low density granular gas described by the
(inelastic) Boltzmann equation, and it is given by \cite{BDKyS98}
\begin{equation}
\label{1} L_{c}=\frac{(2+d) \Gamma \left( d/2 \right)}{2
\pi^{\frac{d-3}{2}} n \sigma^{(d-1)}}\, \left( \frac{\eta^{*}}{2
\zeta^{*}} \right)^{1/2},
\end{equation}
where $n$ is the number of particles density,
$\zeta^{*}(\alpha)=\eta_{0}\zeta_{hcs}/n k_{B}T_{hcs}$, and
$\eta^{*}(\alpha)=\eta(T_{hcs})/\eta_{0}(T_{hcs})$, with $k_{B}$
being the Boltzmann constant, $\eta$ the shear viscosity of the
granular gas, and $\eta_{0}$ its elastic limit. The explicit
expressions of $\zeta^{*}$ and $\eta^{*}$ are given in
\cite{BDKyS98}. In the elastic limit $\alpha =1$, $\zeta^{*}$
vanishes and $\eta^{*}=1$.The accuracy of this prediction has been
verified by direct Monte Carlo simulation of the Boltzmann
equation \cite{BRyM98}.

Another important consequence of the inelasticity of collisions,
largely unexplored, is the presence of an intrinsic noise in the
macroscopic description of the system. Recently \cite{BGMyR04}, it
has been shown that the global energy of a dilute granular gas in
the HCS exhibits fluctuations and time-correlation properties
which are caused by the energy dissipation in collisions and, at a
mesoscopic level, by the presence of velocity correlations.
Although the correlations increase as the inelasticity increases,
they remain relatively small over all the range of values of
$\alpha$ studied ($\alpha \geq 0.6$). Here it will be shown that
the noise is amplified, and in fact it diverges, at the threshold
of the clustering instability.

We have performed molecular dynamics (MD) simulations of a freely
evolving system of $N$ inelastic hard disks ($d=2$), using the
steady-state method, that is based on an {\em exact} mapping of
the HCS onto a steady state. This property  follows from a change
in the time scale, as discussed in detail in \cite{Lu01}. The
steady representation of the HCS removes the difficulties
associated with the rapid cooling of the fluid leading to
numerical inaccuracies very soon. With this method, the
trajectories can be followed for an arbitrary time. The system
considered has been a square box of size $L$ with periodic
boundary conditions. The density in all the results reported in
the following is $n= 0.02 \sigma^{-2}$, so the values of $L$ and
$N$ were consistently changed in the different simulations. The
above value of the density has been found to be low enough as to
guarantee that the average behavior of the system is accurately
described by the Boltzmann equation, even in the unstable region
of the granular gas \cite{ByR04}. Two different values of the
restitution coefficient have been considered, $\alpha= 0.8$ and
$\alpha=0.9$, for which Eq.\ (\ref{1}) gives $L_{c}\simeq 304
\sigma$ and $L_{c}\simeq 413 \sigma$, respectively. We have
studied the behavior of the global properties of the system when
the critical size is approached from below by increasing the value
of $L$ at constant $\alpha$ (and $n$). In all the simulations, we
have checked that the system was actually in the HCS by monitoring
the local velocity and density fluctuations.

\begin{figure}
\includegraphics[scale=0.5,angle=-90]{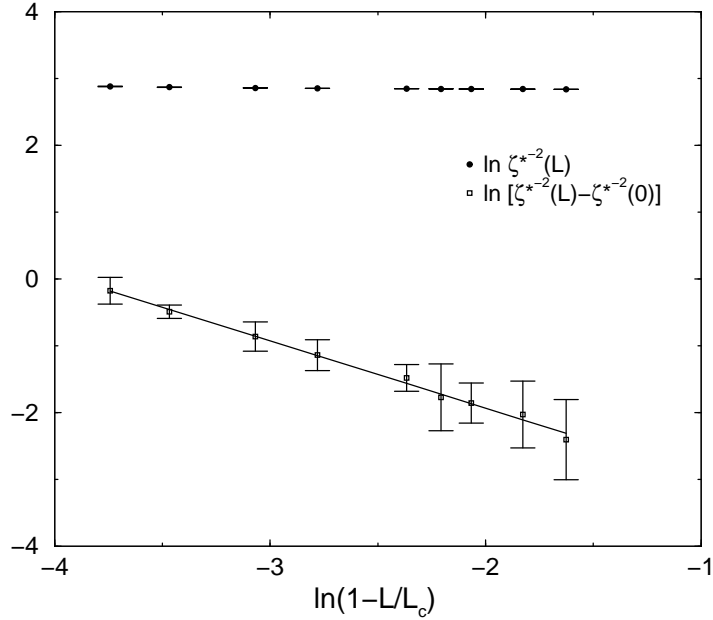}
\caption{Dependence of the scaled cooling rate $\zeta^{*}(L)$ on
the ``distance'' to the instability point $\widetilde{\delta L}
\equiv 1-L/L_{c}$ for a system of inelastic hard disks with
$\alpha=0.9$. The solid line is a fit to the power-law behavior
discussed in the text. \label{f1} }
\end{figure}

The first quantity we have investigated is the reduced cooling
rate $\zeta^{*}$ as a function of the linear size of the system,
at constant $\alpha$. The simulations show that it monotonically
decreases very slowly as $L$ approaches its critical value. The
decay is imperceptible on the scale used in Fig. \ref{f1}, where
$\ln \zeta^{*-2}$ is plotted as a function of $\widetilde{\delta
L} \equiv (L_{c}-L)/L_{c}$ for $\alpha=0.9$. Here and in the
following, the theoretical prediction for $L_{c}$ given by Eq.\
(\ref{1}) has been used. To make the above dependence on $L$
explicit, we have also plotted $\ln [ \zeta^{*-2}(L)-\zeta^{* -2}
(0)]$, where $\zeta^{*}(0)$ is the constant asymptotic value of
the reduced cooling rate far below the transition point, also
obtained from the simulations. The results indicate a behavior of
the form $\zeta^{*-2}(L) \sim \zeta^{*-2}(0) + A_{\zeta}
\widetilde{\delta L}^{-1} $, with $A_{\zeta} \simeq 0.02 $.
Similar results were found for $\alpha=0.8$. If this behavior
persists until the instability point, it would imply that the
cooling rate tends to vanish as $(\widetilde{\delta
L}/A_{\zeta})^{1/2}$. Nevertheless, since in the observed region
$\zeta^{*}(L)$ is dominated by its asymptotic part $\zeta^{*}(0)$,
it is difficult to make any definite statement on this point.

\begin{figure}
\includegraphics[scale=0.5,angle=-90]{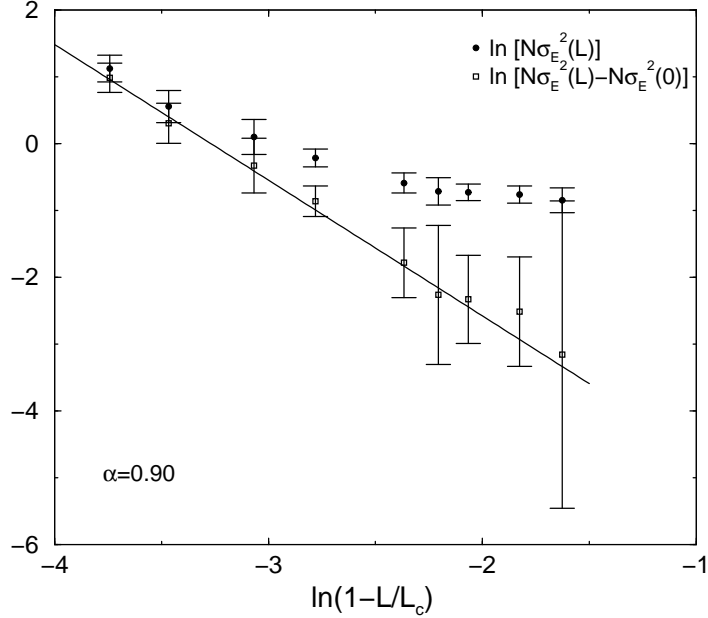}
\caption{Scaled second moment of the total energy fluctuations
$\sigma_{E}^{2}$ as a function of the ``distance'' $
\widetilde{\delta L}$ to the instability point for $\alpha=0.9$.
Note that the second moment has been multiplied by the number of
particles $N$. The solid line is the fit to the power law defined
in the text. \label{f2} }
\end{figure}

Next, let us consider the second moment of the fluctuations of the
total energy $E$,
\begin{equation}
\label{2} \sigma_{E}^{2} \equiv \frac{ \langle E^{2}(t) \rangle -
\langle E(t) \rangle^{2}}{\langle E(t) \rangle^{2}},
\end{equation}
where the angular brackets denote average over the ensemble
generated by a number of trajectories \cite{BGMyR04}. The above
quantity does not depend on time in the HCS, due to the scaling
property of its distribution function. The results, again for
$\alpha=0.9$, are given in Fig.\ \ref{f2}. For $L \ll L_{c}$, $N
\sigma_{E}^{2}(L)$ is practically constant, independent of $L$,
indicating that $\sigma_{E} \propto N^{-1/2}$, as expected. In
fact, its value in this region, that we will denote by
$N\sigma_{E}^{2}(0)$, is accurately predicted by the result
derived in ref. \cite{BGMyR04} by using kinetic theory methods.
Although it is a function of $\alpha$, increasing as $\alpha$
decreases, it remains rather small, at least for not too inelastic
systems. On the other hand, when $L$ approaches its critical
value, the energy fluctuations grow very fast. This is clearer
seen by studying the quantity $N
\left[\sigma_{E}^{2}(L)-\sigma_{E}^{2}(0) \right]$, which is also
plotted in the same figure. It follows that close enough to the
instability point, the second moment of the total energy
fluctuations is accurately described by the critical law
\begin{equation}
\label{3} \sigma_{E}(L) \sim A_{E} (\widetilde{\delta L})^{-1},
\end{equation}
where $A_{E} \simeq 6 \times 10^{-4}$ is a critical amplitude.
Upon writing the above equation, we have used that near the
instability point we can approximate $N \sim N_{c} = nL_{c}^{2}$.
The same behavior is found for $\alpha=0.8$, but with the
amplitude $A_{E}\simeq 1.5 \times 10^{-3}$. Therefore, the
simulation results indicate a divergent critical behavior of the
fluctuations, with a critical exponent $-1$ and an amplitude that
depends on the value of the coefficient of restitution $\alpha$.
Of course, the validity of the above relies on the assumption that
the observed behavior remains the same until the system is
asymptotically close to the instability.

\begin{figure}
\includegraphics[scale=0.5,angle=-90]{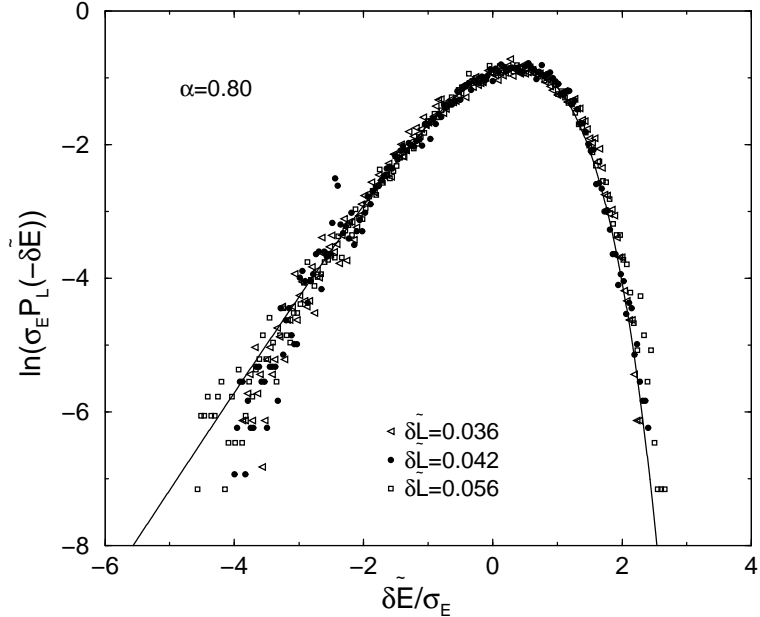}
\caption{Probability density function of the relative total energy
fluctuations $\sigma_{E} P_{L}(-\widetilde{\delta E})$ for a
system of inelastic hard disks with $\alpha=0.8$. The symbols are
from MD simulations and the solid line is Eq. (\ref{7}).
\label{f3} }
\end{figure}

\begin{figure}
\includegraphics[scale=0.5,angle=-90]{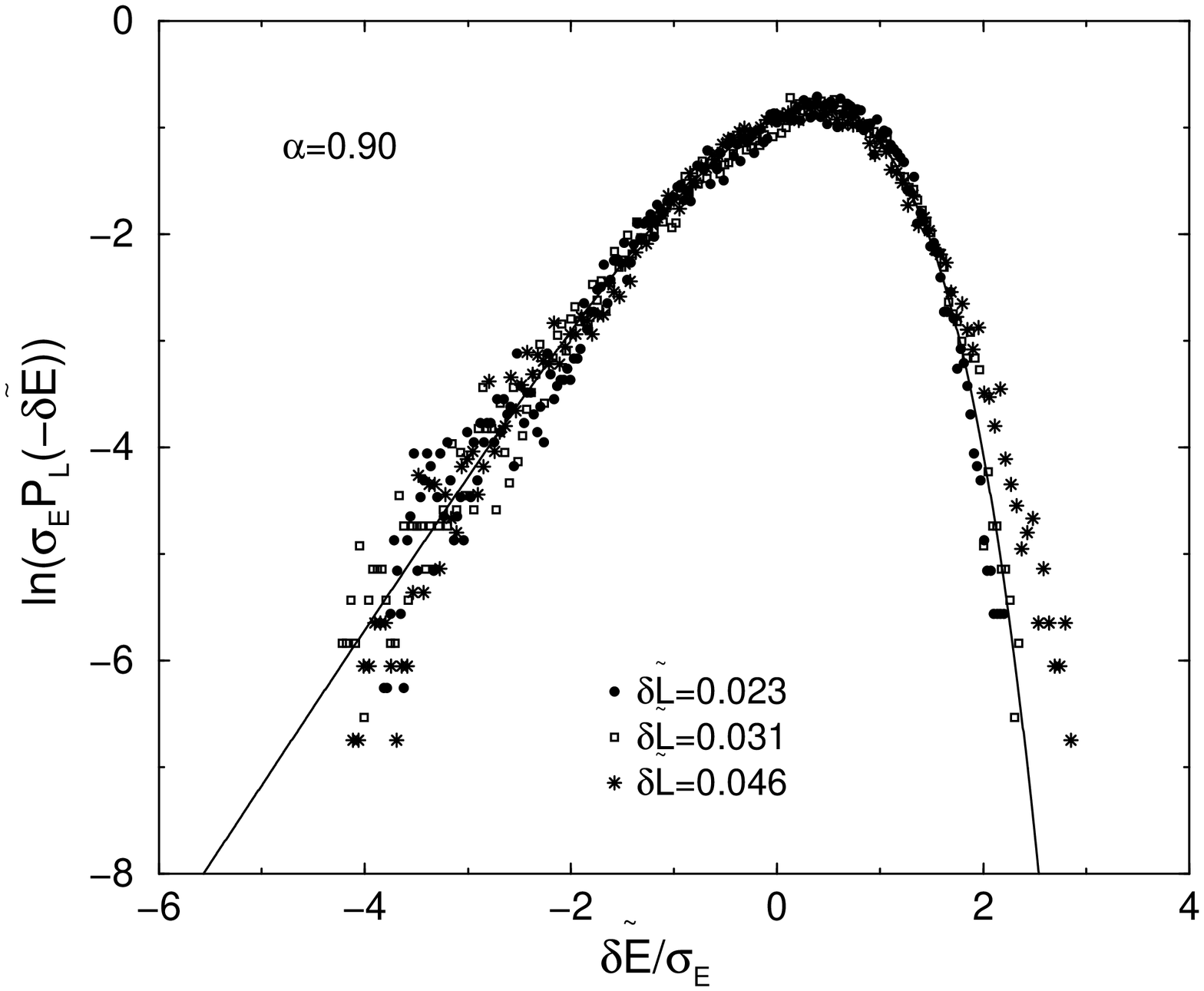}
\caption{The same as in Fig. \ref{f3}, but for $\alpha=0.9$.
\label{f4}}
\end{figure}

To analyze in more detail the structure of the energy
fluctuations, we make the unjustified assumption ({\em a priori})
that the probability distribution function (PDF)
$P_{L}(\widetilde{\delta E})$ of the relative energy fluctuations
$\widetilde{\delta E} \equiv (E(t)-\langle E(t) \rangle)/\langle
E(t) \rangle $ in the vicinity of the clustering instability
verifies the scaling relation
\begin{equation}
\label{4} P_{L}(\widetilde{\delta E})= \frac{1}{\sigma_{E}}
\tilde{f} \left( \frac{\widetilde{\delta E}}{\sigma_{E}} \right),
\end{equation}
where $\tilde{f}$ is a scaling function. It is moreover assumed
that all the dependence of $\tilde{f}$ on $\alpha$ and $L$ occurs
through $\sigma_{E}$. Consider then the function
\begin{equation}
\label{5} \sigma_{E} P_{L} (\widetilde{\delta E}) = \tilde{f}
\left( \frac{\widetilde{\delta E}}{ \sigma_{E}} \right).
\end{equation}
All the $L$ and $\alpha$ dependence has been eliminated on the
right hand side of the above expression when considered as a
function of $\widetilde{\delta E} / \sigma_{E}$. Thus as long as
the assumed scaling holds near the clustering instability, the
data for systems with different sizes should tend to collapse onto
each other as $L$ approaches $L_{c}$. Moreover, the collapse must
occur on a function that does not depend on $\alpha$. To check
this strong prediction, we have computed the PDF of the
fluctuations in the MD simulations. For each trajectory of a
system, the trace of $\widetilde{\delta E}$ has been partitioned
into nonoverlapping bins and the frequency distribution has been
built up using all the trajectories corresponding to the same
values of $L$ and $\alpha$. The resulting normalized distributions
are plotted in Figs. \ref{f3} and $\ref{f4}$, for $\alpha=0.8$ and
$\alpha=0.9$, respectively. In each case, results for different
system sizes are given. Notice that, for reasons that will be
explained below, what has been actually plotted is $\sigma_{E}
P_{L}(-\widetilde{\delta E})$. The data are seen to fall quite
closely on top of each other, as implied by the scaling law
(\ref{5}), over more than three orders of magnitude, specially
when attention is restricted to the results for the two largest
systems. Moreover, the fluctuations are highly non-Gaussian and
asymmetric around the mean value. This is a manifestation of the
presence of a correlation length of the order of the size of the
system, so that it can not be divided into statistical independent
mesoscopic regions and, therefore, there is no reason to expect
the fluctuations of global quantities to be Gaussian.

The shape of PDF's in Figs. \ref{f3} and \ref{f4} looks quite
similar to the functional form found in other equilibrium and
non-equilibrium systems \cite{BHyP98,BChFyal00}. To make this
statement more precise, we have considered the approximated form
of the PDF for the fluctuations of the global magnetization in the
two-dimensional XY model in the harmonic approximation, for
temperatures well below the Kosterlitz-Thouless transition
\cite{BChFyal00},
\begin{equation}
\label{7} \Pi(y)= K \left( e^{x-e^{x}} \right)^{a}, \quad
x=b(y-s), \quad a=\pi/2.
\end{equation}
The high $y$ region of this function is approximately Gaussian,
while it presents an exponential tail for large negative values. A
quite large number of non-equilibrium systems exhibiting
self-organized criticality has been found to show data collapse
with a PDF very similar to $\Pi(y)$ \cite{BChFyal00}. Quite
remarkably, the same kind of behavior had been  previously
observed in experiments with confined turbulent flows
\cite{PHyL99}. In the present context, in principle it should be
$\Pi (y) = \sigma_{E} P_{L}(y)$ with $y=\widetilde{\delta E}$, but
in order to get a good agrement with the MD results, we have to
change $\widetilde{\delta E}$ by $-\widetilde{\delta E}$ in the
identification of $y$. The values of the three parameters in Eq.\
(\ref{7}) follow from the normalization, the zero mean, and unit
variance conditions, with the results $K=2.14$, $b=0.938$, and
$s=0.374$. Therefore, Eq.\ (\ref{7}) has no fitting parameters.
The function $\Pi(y)$ has been also plotted in Figs.\ \ref{3} and
\ref{4}, and a surprisingly good agreement with the simulation
data is found. The change in the sign of the fluctuations, i.e.
the fact that the energy fluctuations in a granular gas are
described by a function that is the symmetric with respect to the
origin of that for ordinary systems, may be due to the dissipative
character of granular systems. In the HCS the gas is continuously
dissipating energy, while in molecular systems energy must be
continuously supplied in order to keep them in a nonequilibrium
steady state. Let us mention that large fluctuations around the
threshold of a symmetry breaking instability have also been
observed recently in granular systems driven by a thermal sidewall
at zero gravity \cite{MPSyS04}. To our knowledge, whether these
fluctuations can be scaled in a similar way to the one described
here has not been investigated yet.

The natural question prompted for the results reported in this
work is whether the agreement of the fluctuation spectra of such a
variety of systems is telling us something about the intrinsic
behavior of a quite general class of systems. If this is the case,
how this class can be characterized {\em a priori}, and the
divergent behavior (\ref{3}) and the specific scaling form
(\ref{7}), or another one very close to it, can be derived on the
ground of general arguments. Specifically in the context of
granular systems, how can the above results be obtained starting
from a microscopic description of the system? Are they captured by
a fluctuating hydrodynamics theory? Is there an underlying
hyperscaling in the sense discussed in \cite{AyG01}? Finally, it
must be stressed that we have restricted ourselves here to
two-dimensional low-density granular gases. Although it seems that
the same kind of results can be expected at higher densities and
also for three-dimensional systems (perhaps with a different
critical exponent), this is something to be verified. These points
are presently under study and some results will be reported
elsewhere.

This research was supported by the Ministerio de Ciencia y
Tecnolog\'{\i}a (Spain) through Grant No. BFM2002-00307 (partially
financed by FEDER funds).

\end{document}